\documentclass[twocolumn,showpacs,superscriptaddress,amsmath,amssymb,prl]{revtex4}
\usepackage{mathrsfs}

\usepackage{graphicx,color}
\usepackage{dcolumn}
\usepackage{bm}
\usepackage{CJK}
\usepackage{epstopdf}

\begin{document}
\title{Predictions of nuclear $\beta$-decay half-lives with machine learning and their impacts on $r$ process}

\author{Z. M. Niu}
\affiliation{School of Physics and Materials Science, Anhui University,
             Hefei 230601, China}
\affiliation{Institute of Physical Science and Information Technology, Anhui University,
             Hefei 230601, China}

\author{H. Z. Liang}
\email{haozhao.liang@riken.jp}
\affiliation{RIKEN Nishina Center, Wako 351-0198, Japan}
\affiliation{Department of Physics, Graduate School of Science, The University of Tokyo,
  Tokyo 113-0033, Japan}

\author{B. H. Sun}
\affiliation{School of Physics and Nuclear Energy Engineering, Beihang University,
             Beijing 100191, China}

\author{W. H. Long}
\affiliation{School of Nuclear Science and Technology, Lanzhou University,
             Lanzhou 730000, China}

\author{Y. F. Niu}
\affiliation{School of Nuclear Science and Technology, Lanzhou University,
             Lanzhou 730000, China}
\affiliation{ELI-NP, ``Horia Hulubei'' National Institute for Physics and Nuclear Engineering,
             30 Reactorului Street, RO-077125, Bucharest-Magurele, Romania}

\date{\today}

\begin{abstract}
Nuclear $\beta$ decay is a key process to understand the origin of heavy elements in the universe, while the accuracy is far from satisfactory for the predictions of $\beta$-decay half-lives by nuclear models up to date. In this letter, we pave a novel way to accurately predict $\beta$-decay half-lives with the machine-learning based on the Bayesian neural network, in which the known physics has been explicitly embedded, including the ones described by the Fermi theory of $\beta$ decay, and the dependence of half-lives on pairing correlations and decay energies. The other potential physics, which is not clear or even missing in nuclear models nowadays, will be learned by the Bayesian neural network. The results well reproduce the experimental data with a very high accuracy and further provide reasonable uncertainty evaluations in half-life predictions. These accurate predictions for half-lives with uncertainties are essential for the $r$-process simulations.
\end{abstract}


\maketitle

The origin of heavy elements, e.g., how and where the rare elements such as gold and platinum were created in the Universe, is a fascinating but still unanswered question of physics~\cite{Haseltine2002Discover}. It relates to many branches of science, notably astrophysics and nuclear physics~\cite{Burbidge1957RMP}, so the answer to this question necessitates joint efforts of scientists from various fields. The rapid neutron-capture process ($r$ process) is responsible for producing about half of the elements heavier than iron (Fe), in fact the only mechanism for producing elements beyond Bi. However, the understanding of $r$ process still remains many mysteries from the points of view of nuclear physics and astrophysics. Recent multi-messenger observations including the gravitational-wave signal and multi-wavelength electromagnetic counterparts strongly support neutron-star merger to be a site of production of heavy elements via $r$ process~\cite{Abbott2017PRL, Arcavi2017Nature, Pian2017Nature}. On the other hand, the measurements of nuclear properties also achieved great progress with the development of radioactive ion beam (RIB) facilities, especially the region around $N=82$~\cite{Wang2017CPC, Audi2017CPC, Franzke2008MSR, Sun2015FP, Nishimura2011PRL, Lorusso2015PRL, Wu2017PRL}. These new observations and measurements make the $r$ process become a very hot topic in physics nowadays.

The $r$ process which accounts for the origin of many heavy elements involves many unstable neutron-rich nuclei, namely the exotic ones. The experimental measurements are approaching the $r$-process pathes around $N=82$, while still far away from ones around $N=126$. Consequently, the reliable theoretical predictions of nuclear properties are necessary to the $r$-process simulations. Nuclear $\beta$ decay is a decay process of nucleus by emitting an electron and a neutrino, and hence it generates the new element with the proton number larger than parent nucleus. Such that nuclear $\beta$ decay governs the abundance flow between neighboring isotopic chains in the $r$ process, and plays a key role in understanding the origin of heavy elements. However, the predictions of nuclear $\beta$-decay half-lives are rather difficult for nuclear theory due to the complexity in both interactions and nuclear many-body calculations. The accurate predictions of nuclear $\beta$-decay half-lives still remain an important but unsolved problem in nuclear physics. Theoretically, massive efforts have been devoted to this topic by developing the nuclear models based on various approximations or in a limited configuration space, such as gross theory (GT)~\cite{Takahashi1969PTP, Tachibana1990PTP, Nakata1997NPA, Koura2017PRC}, quasiparticle random phase approximation (QRPA) approach~\cite{Engel1999PRC, Minato2013PRL, Niu2013PLB, Moller2003PRC, Marketin2016PRC, Borzov2000PRC}, and shell model~\cite{Langanke2003RMP, Pinedo1999PRL, Suzuki2012PRC, Zhi2013PRC}. Unfortunately, the evaluations of theoretical uncertainties of $\beta$-decay half-lives are still very scarce in literatures, although they are essential to understand the reliability of theoretical predictions and further their impacts on the $r$ process~\cite{Mumpower2016PPNP}.

The machine learning has been widely applied in engineering, such as the pattern recognition and classification tasks. It is very powerful in extracting pertinent features for complex non-linear systems with complicated correlations, which are hard or even impossible to be tackled by traditional models. Therefore, it also provides a powerful tool in physics research, including particle physics~\cite{Baldi2014NC, Pang2014NC, Brehmer2018PRL} and condensed matter physics~\cite{Carrasquilla2017NaturePhysics, Carleo2017Science}. In nuclear physics, the machine learning has been introduced to predict some nuclear properties based on the traditional neural network~\cite{Costiris2009PRC, Zhang2017JPG}. Comparing with conventional nuclear models, it constructs a neural network complex enough to predict nuclear properties accurately while with many parameters, which in general accompanies with the over-fitting problem and undetermined theoretical uncertainties. In contrast to that, the machine learning with Bayesian neural network (BNN) approach~\cite{Bishop2006Book, Neal1996Book, Utama2016PRC, Niu2018PLB} can avoid the over-fitting automatically by including prior distribution and can quantify the uncertainties in its predictions naturally. Therefore, it is quite promising to predict nuclear $\beta$-decay half-lives accurately and give reasonable uncertainty evaluations with the BNN approach.

To better predict nuclear properties, it is important to include those well known physics as much as possible before applying BNN approach. Therefore, a more effective strategy is to use the BNN approach to improve the predictions of nuclear models. In this work, we will first propose a theoretical formula to predict nuclear $\beta$-decay half-lives based on the Fermi theory, and the BNN approach is then employed to improve the predictions of $\beta$-decay half-lives. Their impacts on solar $r$-process simulations are investigated as well.

Let us start with the well-known Fermi theory of $\beta$ decay~\cite{Fermi1934ZP}, in which nuclear $\beta$-decay half-life in the allowed Gamow-Teller approximation is predicted by
\begin{eqnarray}\label{Eq:T12Fermi}
    T_{1/2}
  =\frac{D}{g_A^2 \sum_{E_m<Q_\beta} B(E_m) f(Z,A,E_m)},
\end{eqnarray}
where $D=6163.4$~s, and $g_A$ is the effective weak axial nucleon coupling constant. $B(E_m)$ is the transition strength from the ground state of parent nucleus to the excited state $m$ of the daughter nucleus, as a function of transition energy $E_m$. The total $\beta$-decay energy $Q_\beta = m_P - m_D - m_e$, where $m_P$, $m_D$ and $m_e$ are the masses of parent nucleus, daughter nucleus, and electron, respectively. $f(Z,A,E_m)$ is the integrated lepton $({\rm e}^{-}, \bar{\nu}_{\rm e})$ phase volume, where Coulomb screening and relativistic nuclear finite-size corrections have been considered~\cite{Niu2013PLB}.

When $E_m\gg m_e$, nuclear half-lives are mainly determined by $f(Z,A,E_m)$ since it is proportional to $E_m^5$. However, the accurate predictions of $B(E_m)$ are still very difficult for present nuclear models, which can only reproduce experimental half-lives within a few orders of magnitude. Therefore, we could approximatively predict nuclear half-life with
\begin{eqnarray}\label{Eq:T12afQd}
    T_{1/2} = a/f(Z,A,E_m),
\end{eqnarray}
where $E_m$ is estimated by $E_m=Q_\beta-b(1-\delta)/\sqrt{A}$ with $\delta=1,0,-1$ for even-even nuclei, odd $A$ nuclei, and odd-odd nuclei, respectively. In this work, the $Q_\beta$ are calculated using the mass predictions of WS4 model~\cite{Wang2014PLB}. The parameters $a=4.96$ and $b=8.51$ are determined by the best fitting to experimental $\beta$-decay half-lives from NUBASE2016~\cite{Audi2017CPC}.

\begin{figure}
\includegraphics[width=7.3cm]{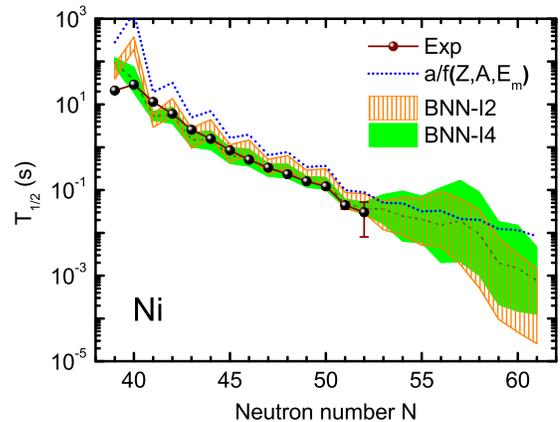}
\caption{(Color online) $\beta$-decay half-lives of Ni isotopes. The experimental values in NUBASE2016 are denoted by spheres. The half-life predictions with Eq.~(\ref{Eq:T12afQd}) are shown by the dotted line, and their counterparts improved by BNN-I2 and BNN-I4 approaches and their uncertainties are shown by vertical line hatched region and green hatched region, respectively. The mean predicted half-lives of BNN-I4 approach are marked by the dashed line.}
\label{Fig:NiDiffBNN}
\end{figure}

Since nuclear half-lives vary by many orders of magnitude, the root-mean-square (rms) deviation of logarithm of half-life $\log_{10}(T_{1/2})$ is usually employed to evaluate the accuracy of nuclear models
\begin{eqnarray}\label{Eq:rmsT12}
    \sigma_{\rm rms}(\log_{10}T_{1/2}) = \sqrt{
                   \frac{\sum\limits_{i=1}^n \left[\log_{10}(T_{1/2}^{\rm exp}/T_{1/2}^{\rm th})\right]_i^2}{n}
                   },
\end{eqnarray}
where $T_{1/2}^{\rm exp}$ and $T_{1/2}^{\rm th}$ are the experimental and theoretical half-lives, respectively, and $n$ is the number of nuclei in a given evaluation set. In this work, the experimental data are taken from NUBASE2016, in which only those nuclei with $T_{1/2}<10^6$ seconds and $Z, N\geqslant 8$ are adopted. It is surprising that this ``oversimplified" formula can already reproduce the known half-lives with $\sigma_{\rm rms}(\log_{10}T_{1/2})=0.81$. This is even similar to the precision obtained by the sophisticated QRPA model based on the finite-range droplet model (FRDM)~\cite{Moller2003PRC}, whose $\sigma_{\rm rms}(\log_{10}T_{1/2})$ is $0.82$. Therefore, we believe Eq.~(\ref{Eq:T12afQd}) must grasp the main physics in half-life predictions. Taking Ni isotopes as the typical examples, it is shown in Fig. \ref{Fig:NiDiffBNN} that Eq. (\ref{Eq:T12afQd}) generally reproduces the data within less than one order of magnitude of accuracy, while showing systematic overestimation of nuclear $\beta$-decay half-lives with too strong odd-even staggering. Such discrepancies from the data, which account for the physics missing in nuclear models, will be dealt with the BNN approach in this work, and furthermore, the resulting uncertainties of half-life predictions can be estimated as well. For the details of BNN approach, please refer to the Appendix.

Figure~\ref{Fig:NiDiffBNN} shows the predictions of BNN-I2 and BNN-I4 approaches for Ni isotopes, in comparison with the ones given by Eq.~(\ref{Eq:T12afQd}) and the data. It is found that the BNN-I2 approach can eliminate the systematic overestimation of half-lives in the predictions of Eq.~(\ref{Eq:T12afQd}), while its odd-even staggering is still remained. Implemented with the BNN-I4 approach, this odd-even staggering is removed in a large extent and the resulting predictions are in excellent agreement with the experimental data. It also demonstrates that the BNN approach, including the known physics, paves an effective way for the reliable and accurate prediction of nuclear $\beta$-decay half-life. In the following, therefore, we will only show our results based on the BNN-I4 approach.

\begin{figure}
\includegraphics[width=7.3cm]{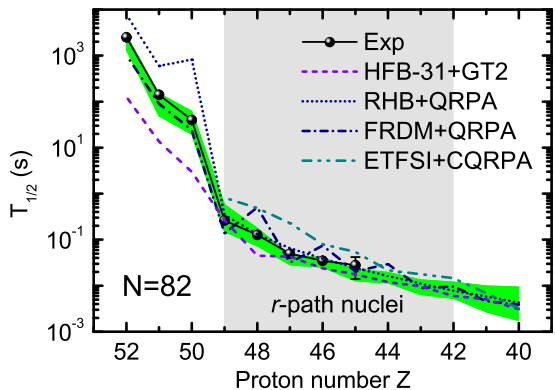}
\caption{(Color online) $\beta$-decay half-lives of Ni isotopes and $N=82$ isotones. The predictions of BNN-I4 approach are shown by the green hatched regions. The experimental values in NUBASE2016 are denoted by spheres. For comparison, the theoretical results from HFB-31~+~GT2, RHB~+~QRPA, FRDM~+~QRPA, and ETFSI~+~CQRPA models are shown by the dashed, dotted, dash-dotted, and dash-dot-dotted lines, respectively.}
\label{Fig:T12Comps}
\end{figure}

As further illustration, Figure~\ref{Fig:T12Comps}, taking $N=82$ as examples, shows the comparison between the predictions of BNN-I4 approach and other successful theoretical models, including Hartree-Fock-Bogoliubov (HFB-31)~+~GT2~\cite{Tachibana1990PTP}, relativistic Hatree-Bogoliubov (RHB)~+~QRPA~\cite{Marketin2016PRC}, FRDM~+~QRPA~\cite{Moller2003PRC}, and extended Thomas-Fermi plus Strutinsky integral (ETFSI)~+~continuum QRPA (CQRPA)~\cite{Borzov2000PRC} models. Once again, the results of BNN-I4 approach are in rather good agreement with the experimental data, even completely agree with the experimental data within uncertainties. On the contrary, the results from other theoretical models generally show large deviations from the experimental data. When extrapolated to the unknown region, the uncertainties of BNN-I4 predictions increase slightly for $N=82$ isotones, while they increase remarkably for Ni isotopes (see Fig.~\ref{Fig:NiDiffBNN}). It is interesting to notice that the BNN-I4 half-life predictions of Ni isotopes slowly decrease in the region $N=51 \sim 58$ and suddenly drop at $N=59$. This phenomenon may originate from the microscopic shell effect, since the last occupied single-neutron orbitals are all $1g_{7/2}$ for nuclei $^{79\text{--}86}$Ni as indicated by the calculations with the mean-field model. Since the uncertainties of BNN predictions in this region are large, future measurements on the half-lives of Ni isotopes are necessary to confirm whether or not this phenomenon is real.

\begin{figure}
\includegraphics[width=7.3cm]{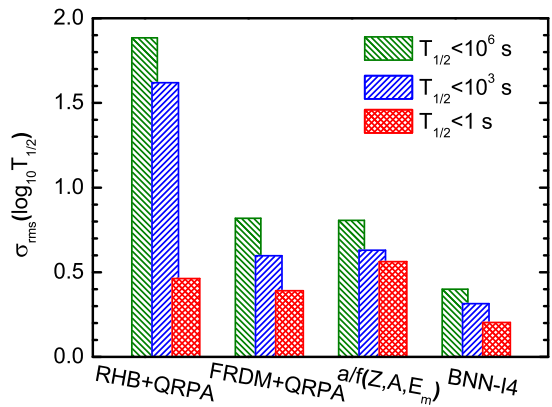}
\caption{(Color online) The rms deviations $\sigma_{\rm rms}(\log_{10}T_{1/2})$ with respect to the known $\beta$-decay half-lives from NUBASE2016. Three sets of nuclei with $T_{1/2}<10^6$ s, $T_{1/2}<10^3$ s, and $T_{1/2}<1$ s are used in the rms evaluations.}
\label{Fig:rmsT12}
\end{figure}

In order to evaluate the global reliability of BNN approach to predict nuclear $\beta$-decay half-lives, the rms deviations $\sigma_{\rm rms}(\log_{10}T_{1/2})$ of BNN-I4 predictions with respect to the experimental data are presented in Fig.~\ref{Fig:rmsT12}. For comparison, the corresponding results based on RHB~+~QRPA, FRDM~+~QRPA, and Eq.~(\ref{Eq:T12afQd}) are shown as well. It can be clearly seen that the theoretical approaches in general better reproduce the experimental data of nuclei with shorter half-lives. The BNN-I4 approach significantly improve the half-life predictions of Eq.~(\ref{Eq:T12afQd}) with the accuracy much better than the selected models, particularly for nuclei with half-lives shorter than $1$~s. It is worthwhile to mention that the nuclei along or near the $r$-process path are in general characterized by the typical half-lives less than $1$~s. For these nuclei, which are in particular our focus, the $\sigma_{\rm rms}(\log_{10}T_{1/2})$ of BNN-I4 approach is only $0.20$. Namely the BNN-I4 approach can describe these relevant nuclear half-lives within a factor of two with respect to the experimental data ($10^{0.20}=1.58$). Such high accuracy, which is achieved for the first time, is essential for the $r$-process simulations.

\begin{figure}
\includegraphics[width=7.3cm]{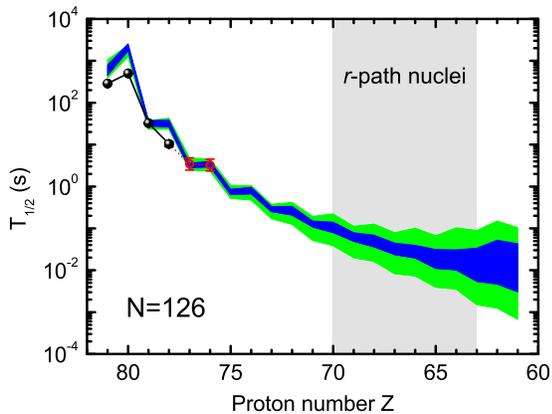}
\caption{(Color online) $\beta$-decay half-lives of $N=126$ isotones. The green hatched region shows the predictions of BNN-I4 approach, whose learning data are only taken from NUBASE2016. The blue hatched region shows results of BNN-I4 approach as well, while its learning data are extended to include three extra $\beta$-decay half-lives for each isotope (denoted by open circles) towards neutron-drip line. The experimental values in NUBASE2016 are denoted by spheres.}
\label{Fig:Neq126Err}
\end{figure}

Furthermore, with the present Bayesian scheme, it is natural to ask ``\textit{In case a few more nuclear half-lives are determined towards the neutron-drip line, how can these new data affect the predictions?}'' This is in particular an interesting question, since it is foreseen that even with the next generation RIB facilities it is still not feasible to reach all the $r$-process-path nuclei experimentally. Here, let us assume three more new $\beta$-decay half-lives were measured towards neutron-drip line for each isotope, which are taken as the new experimental data in the BNN-I4 prediction. By further including these new artificial data to the learning set, the BNN-I4 approach is trained again. The resulting predictions are shown in Fig.~\ref{Fig:Neq126Err} by taking $N=126$ isotones as examples, in which the original BNN-I4 results with learning data only from NUBASE2016 are denoted with the green hatched region and the blue one corresponds to the BNN-I4 results with three more artificial learning data. As expected, in the known region, even if the new artificial data are included, the uncertainties of BNN-I4 predictions remain almost the same as before. However, when extending the unknown region, the new artificial data make the uncertainties decrease by about a factor of $3$. It is very important to the $r$-process studies. Although many $r$-process-path nuclei around $N=126$ are still hard to be measured even in the new-generation RIB facilities, this fact tells us the uncertainties of half-life predictions can be significantly reduced with only a few more measured data towards the neutron-drip line for each isotope.

\begin{figure}
\includegraphics[width=7.3cm]{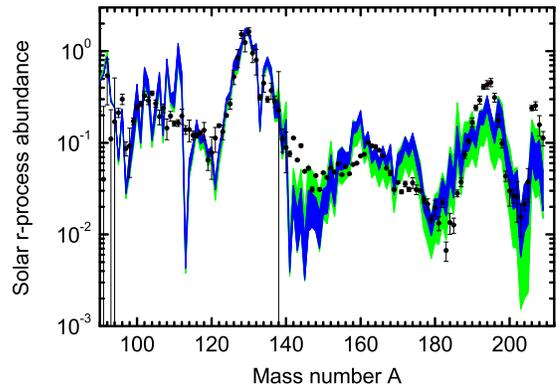}
\caption{(Color online) Impact of nuclear $\beta$-decay half-lives on solar $r$-process calculations. The uncertainties bands for solar $r$-process abundances are due to the uncertainties of nuclear $\beta$-decay half-lives. The green and blue bands, respectively, correspond to the calculations using BNN-I4 half-life predictions, whose learning data are whether to include three new artificial $\beta$-decay half-lives for each isotope or not. The solar $r$-process abundances~\cite{Sneden2008ARAA} are denoted by filled circles.}
\label{Fig:RpAbund}
\end{figure}

Nuclear $\beta$-decay governs the $r$-process abundance flow between neighboring isotopic chains, so the uncertainties in nuclear $\beta$-decay half-lives would affect the $r$-process abundance distributions. Figure~\ref{Fig:RpAbund} presents the solar $r$-process calculations based on the classical $r$-process model~\cite{Pfeiffer1997ZPA, Sun2008PRC, Niu2009PRC}. In our calculations, the experimental data including nuclear masses~\cite{Wang2017CPC} and $\beta$-decay half-lives~\cite{Audi2017CPC} are used if available, otherwise we employed the WS4 model to determine the unknown masses and the BNN-I4 approach for the half-lives. The uncertainties bands in Fig.~\ref{Fig:RpAbund} for the solar $r$-process abundances are due to the uncertainties of nuclear $\beta$-decay half-lives, which come from the experimental errors if available, otherwise come from the uncertainties estimated with BNN-I4 approach. The green and blue bands correspond to the results without and with three new artificial data for each isotope when training the BNN-I4 half-life predictions, respectively.

Notice that the measurements have approached the $r$-process path at $N=82$, while they are still far away from the ones around $N=126$. As a result, the half-life uncertainties for $r$-process-path nuclei around $N=82$ are much smaller than those for $r$-process-path nuclei around $N=126$, see Figs.~\ref{Fig:T12Comps} and \ref{Fig:Neq126Err}. Coincidentally, as shown Fig.~\ref{Fig:RpAbund}, the uncertainties of solar $r$-process calculations at $A\gtrsim 140$ are significantly larger than those at $A\lesssim 140$. However, as indicated by the theoretical uncertainties with BNN-I4 approaches, it is quite expectable that the large uncertainties for the $r$-process abundances at $A\gtrsim 140$ can be remarkably reduced with several new data towards the neutron-drip line measured. It also indicates that future relevant experiments would significantly improve our understanding on the $r$ process.

In summary, the machine-learning approach based on the Bayesian neural network is employed to predict nuclear $\beta$-decay half-lives accurately and give reasonable uncertainty evaluations. To possess better predictive power, a theoretical formula for $\beta$-decay half-lives with only two parameters is proposed based on the Fermi theory, which can already reproduce the data in a similar precision as the sophisticated FRDM+QRPA model. Then the BNN approach is trained to improve the predictions of the proposed formula by simulating the missing physics. It is found that after including more physics features related to the pairing effects and the decay energies, the machine learning approach can precisely describe the general evolution of half-lives along isotopic and isotonic chains, including the odd-even effects which are suffering the nuclear models. Collaborating with the predicted nuclear $\beta$-decay half-lives and theoretical uncertainties, the impact on the $r$-process abundance distributions is investigated. It is found that that the uncertainties of $\beta$-decay half-lives have consistently large influence on the solar $r$-process calculations when $A\gtrsim 140$. Fortunately, as revealed by the BNN approaches, the large uncertainties can be remarkably reduced if a few more nuclear half-lives are further determined towards the neutron-drip line for each isotope. It also becomes quite expectable that future measurements on half-lives could substantially improve our understanding on the $r$ process.

\section*{Acknowledgements}
We are grateful to Prof. T. Hatsuda and Dr. K. Yoshida for the fruitful discussions. This work was partly supported by the National Natural Science Foundation of China under Grants No.~11875070, No. 11675065, and No.~11711540016, the Natural Science Foundation of Anhui Province under Grant No.~1708085QA10, the Open fund for Discipline Construction, Institute of Physical Science and Information Technology, Anhui University, the JSPS Grant-in-Aid for Early-Career Scientists under Grant No.~18K13549, the JSPS-NSFC Bilateral Program for Joint Research Project on Nuclear mass and life for unravelling mysteries of the $r$ process, and the RIKEN iTHEMS program.

\section*{Appendix}

In the Bayesian approach, the model parameters $\bm{\omega}$ are described probabilistically while are not fixed values as in our traditional view. Suppose we have a set of data $D=\{(\bm{x_1}, t_1), (\bm{x_2}, t_2), ..., (\bm{x_n}, t_n)\}$, where $\bm{x_k}$ and $t_k$ $(k=1, 2, ..., n)$ are input and output data, $n$ is the number of data. Then the probability distribution of $\bm{\omega}$ after the data $D$ are taken into account, posterior distribution $p(\bm{\omega}|D)$, is given based on Bayes' theorem,
\begin{eqnarray}\label{Eq:BNN}
  p(\bm{\omega}|D) = \frac{p(D|\bm{\omega})p(\bm{\omega})}{p(D)} \propto p(D|\bm{\omega})p(\bm{\omega}),
\end{eqnarray}
where $p(\bm{\omega})$ is prior distribution based on our background knowledge, $p(D|\bm{\omega})$ is the likelihood function, and $p(D)$ is a normalization constant, which ensures the posterior distribution is a valid probability density and integrates to one.

\begin{figure}[htbp]
\includegraphics[width=8cm]{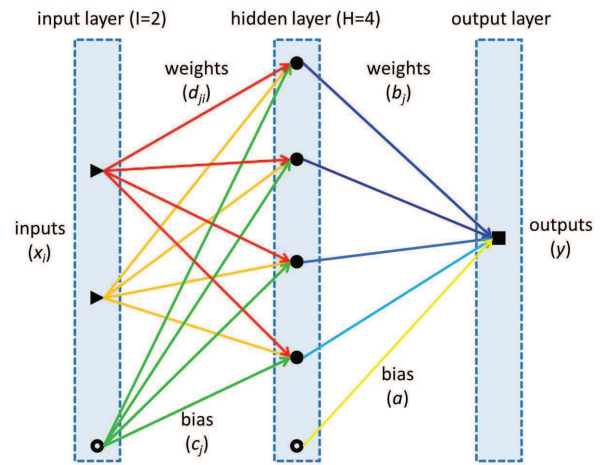}
\caption{(Color online) A schematic diagram for a neural network with a single hidden layer, four neurons ($H=4$), and two input variables ($I=2$).}
\label{Fig:DiagNN}
\end{figure}

In this work, the prior distributions $p(\bm{\omega})$ are set as Gaussian distributions with zero means. The precisions (inverse of variances) of these Gaussian distributions are set as gamma distributions as in Ref.~\cite{Niu2018PLB}, which can make the precisions vary over a large range and the optimal values of precisions are then automatically found during the sampling process. The likelihood function $p(D|\bm{\omega})$ usually employ a Gaussian distribution, $p(D|\bm{\omega})=\exp(\chi^2/2)$, where
\begin{eqnarray}\label{Eq:chiSquare}
  \chi^2=\sum_{n=1}^N\left[\frac{S(\bm{x}; \bm{\omega})-t_n}{\Delta t_n}\right]^2.
\end{eqnarray}
Here, $\Delta t_n$ is the associated noise error, and the inverse of its square $1/\Delta t_n^2$ is set to a gamma distribution as in Ref.~\cite{Niu2018PLB}. The function $S(\bm{x}; \bm{\omega})$ in Bayesian neural network (BNN) approach is a neural network, i.e.
\begin{eqnarray}\label{Eq:NeuralNetwork}
  S(\bm{x}; \bm{\omega})=a+\sum_{j=1}^H b_j \tanh\left(c_j+\sum_{i=1}^I d_{ji} x_i\right),
\end{eqnarray}
where $\bm{x}=\{x_i\}$ and $\bm{\omega}=\{a, b_j, c_j, d_{ji}\}$. $H$ and $I$ are the number of neurons in the hidden layer and the number of input variables, respectively. A schematic diagram for a neural network with a single hidden layer, four neurons ($H=4$), and two input variables ($I=2$) is shown in Fig.~\ref{Fig:DiagNN}.

After specifying the the prior distribution $p(\bm{\omega})$ and likelihood function $p(D|\bm{\omega})$, the posterior distribution $p(D|\bm{\omega})$ can then be obtained by sampling using the Markov chain Monte Carlo algorithm. With the posterior distribution $p(D|\bm{\omega})$, the BNN prediction can be calculated by
\begin{eqnarray}
  \langle S \rangle=\int S(\bm{x}; \bm{\omega}) p(D|\bm{\omega}) d\bm{\omega},
\end{eqnarray}
whose uncertainty is estimated using $\Delta S= \sqrt{\langle S^2 \rangle - \langle S \rangle^2}$.

In this work, the BNN approach is employed to reconstruct residuals between $\log_{10}(T_{1/2}^{\rm exp})$ and $\log(T_{1/2}^{\rm th})$, i.e.,
\begin{eqnarray}
  t_k =\log_{10}(T_{1/2}^{\rm exp}) - \log(T_{1/2}^{\rm th})
      =\log_{10}(T_{1/2}^{\rm exp}/T_{1/2}^{\rm th}).
\end{eqnarray}
The half-life predictions with BNN approaches are then as
\begin{eqnarray}
T_{1/2}^{\rm BNN}=T_{1/2}^{\rm th}\times 10^{S(\bm{x}; \bm{\omega})},
\end{eqnarray}
where $T_{1/2}^{\rm th}$ are calculated with Eq.~(2) in the Letter. Since the $\beta$-decay half-lives are sensitive to the pairing effects and the decay energies, we further introduce $\delta$ and $Q_\beta$ as the inputs of neural network apart from $Z$ and $N$, i.e. $\bm{x}=(Z, N, \delta, Q_\beta)$. For comparison, another neural network with $\bm{x}=(Z, N)$ is also constructed. For simplicity, we will use BNN-I2 and BNN-I4 to denote the BNN approaches with $\bm{x}=(Z, N)$ and $\bm{x}=(Z, N, \delta, Q_\beta)$, respectively. Their numbers of neurons are taken as $H=30$ and $H=20$, respectively. The corresponding neural networks are trained by using $900$ learning data, which are randomly selected from NUBASE2016~\cite{Audi2017CPC}.



\end{document}